\documentclass[twocolumn,superscriptaddress,floatfix]{revtex4}
\tighten
\usepackage{graphicx}

\begin{document}

 \title{ Nonequilibrium $1/f$ Noise in Low-doped Manganite Single Crystals}

\author{X. D. Wu}
 \affiliation{Department of Physics, Ben Gurion University of the Negev, P.O.Box 653, 84105 Beer-Sheva, Israel}
 \affiliation{Department of Materials Engineering, Monash University, Clayton, Australia, 3800}

\author{B. Dolgin}
\affiliation{Department of Physics, Ben Gurion University of the Negev, P.O.Box 653, 84105 Beer-Sheva, Israel}

\author{G. Jung}
\affiliation{Department of Physics, Ben Gurion University of the Negev, P.O.Box 653, 84105 Beer-Sheva, Israel}

\author{V. Markovich}
\affiliation{Department of Physics, Ben Gurion University of the Negev, P.O.Box 653, 84105 Beer-Sheva, Israel}

\author{Y. Yuzhelevski}
 \affiliation{Department of Physics, Ben Gurion University of the Negev, P.O.Box 653, 84105 Beer-Sheva, Israel}

\author{M. Belogolovskii}
 \affiliation{Donetsk Physical and Technical Institute, National Academy of Sciences of Ukraine, 83114, Donetsk and Scientific and Industrial Concern 'Nauka', 04116 Kyiv, Ukraine}

\author{Ya. M. Mukovskii}
 \affiliation{Moscow State Steel and Alloys Institute, 119049, Moscow, Russia}

\date{\today}

\begin{abstract}
$1/f$ noise in current biased La$_{0.82}$Ca$_{0.18}$MnO$_{3}$
crystals has been investigated. The temperature dependence of the
noise follows the resistivity changes with temperature suggesting
that resistivity fluctuations constitute a fixed fraction of the
total resistivity, independently of the dissipation mechanism and
magnetic state of the system. The noise scales as a square of the
current as expected for equilibrium resistivity fluctuations.
However, at 77 K at bias exceeding some threshold, the noise
intensity starts to decrease with increasing bias.  The appearance
of nonequilibrium noise is interpreted in terms of bias dependent
multi-step indirect tunneling.
\end{abstract}

\maketitle

A fundamental interest in mixed-valence manganese perovskites arises
from their strongly spin-dependent conductivity and pronounced
manifestations of phase separation (PS). In a complex and rich phase
diagram of La$_{1-x}$Ca$_{x}$MnO$_3$ (LCMO) manganites the critical
doping level $x_C=0.225$ separates ferromagnetic (FM) insulating
ground state at $x< x_C$ from FM metallic ground state above $x_C$.
In the doping range $0.17<$x$<0.25$ a mixed FM state composed of
insulating and metallic FM phases with different levels of orbital
ordering appears below Curie temperature $T_C$.\cite{papa}
Therefore, physical mechanisms dominating transport properties of
low-doped LCMO, $x\leq x_C$, became remarkably different when the
temperature is changing. Hopping conductivity controls transport in
the paramagnetic (PM) insulating regime at $T>T_C$. Intrinsic PS
associated with metal-insulator (M-I) transition at $T~T_C$ leads to
percolation conductivity in the FM state at $T< T_C$. Low
temperature resistivity is likely dominated by tunneling across
intrinsic barriers associated with extended structural defects, such
as twins and grain boundaries, and/or with inclusions of  insulating
FM phase interrupting metallic percolating paths.\cite{prb018}

Noise measurements are known to provide a unique insight into
dynamics of solid state systems. A peak of the $1/f$ noise resulting
from the percolating transition around $T_C$ was found in various
manganite thin films.\cite{podzorov,reutler,ziese} Grain boundary
junctions were identified as sources of $1/f$ noise in manganite
thin films.\cite{gross} Non-gaussian random telegraph fluctuations
were ascribed to PS at low temperatures.\cite{raquet,merithew}

In such complex system as low-doped LCMO one expects that noise data
will allow to get a deeper insight into the dynamics of dissipation
processes associated with different transport mechanisms. In this
paper we report on electric noise properties in dc current biased
low-doped LCMO single crystals which include the appearance of
nonequilibrium $1/f$ noise at low temperatures.

The experiments were performed with La$_{0.82}$Ca$_{0.18}$MnO$_{3}$
single crystals grown by a floating zone method, as described
elsewhere.\cite{prb018} As grown crystals were cut into individual
samples in the form of thin bars with the longest dimension along
the $<110>$ crystalline direction. Current and voltage leads were
attached to pre-evaporated gold contacts by a conducting silver
epoxy bound. Current-voltage characteristics $(I-V)$, differential
resistance $(R_d=dV/dI)$, and noise spectra were measured at various
temperatures using a standard four-point contact arrangement. The
measurements reported here were all performed at zero applied
magnetic field.

For noise measurements the voltage drop across dc current biased
sample was amplified by a home made very low noise preamplifier
placed directly on the cryostat top, and processed by a dynamic
signal analyzer. Instrumental noise originating from the measuring
chain was eliminated by subtracting the reference spectrum, recorded
at zero current flow in the LCMO sample, from the spectrum acquired
with bias current. Possible contact contributions to the noise were
excluded by using a high impedance ballast resistors in series with
dc current source.

The temperature dependence of the resistivity measured under low dc
current bias of 10 $\mu$A is shown in Fig. 1. At $T>T_C$ the sample
is in PM insulating state. At the temperature of metal-to-isolator
transition at $T_{MI}\approx T_C \approx 180$ K, coinciding with PM
to FM transition, the $\rho(T)$ reaches a local
maximum.\cite{prb018} Below $T_{MI}$ the ferromagnetic metallic
state forms percolating conducting paths in the insulating matrix
and the resistance decreases with decreasing temperature. The low
temperature upturn at temperatures below $T_{min}\sim 150$ K can be
ascribed to the increasing influence of intrinsic tunneling.

The noise spectra were recorded at different bias currents at
temperatures of 77 K, 126 K, 150 K (close to $T_{min}$), 175 K, 185
K, 225 K (in the vicinity of expected Jahn-Teller structural
transition) and 290 K. 126 K, 175 K, and 185 K are the temperatures
at which the sample has the same resistance under much different
dissipation mechanisms. A typical experimental voltage noise power
spectral density (PSD) $S_V$ is shown in the inset to Fig. 1. In the
entire experimental range of temperatures and currents we have
observed $S_v \propto 1/f^\alpha$ with  $0.8 < \alpha < 1.1$ and
only at highest currents exceeding 5 mA,  $\alpha$ as high as 1.5
was recorded at $T=77$ K.

\begin{figure}
\includegraphics*[width=\columnwidth]{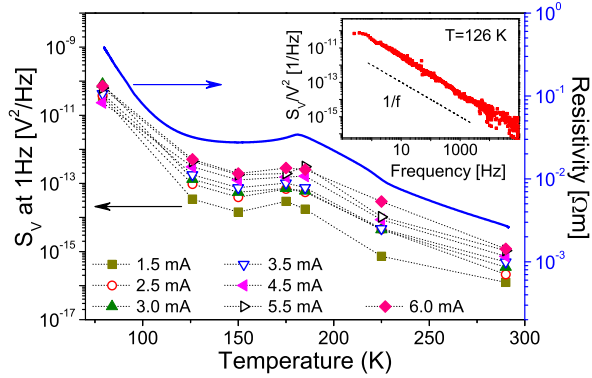}
\caption {Resistiity  and $1/f$ noise intensity at $f=1$ Hz as a
function of temperature. $\rho(T)$ has been recorded using 10 $\mu$A
dc current bias. Currents for noise measurements are specified in
the legend. Inset shows a typical PSD of the noise.} \label{R(T)}
\end{figure}

Temperature dependence of the noise shown in Fig. 1 seems to follow
temperature changes of the resistance, independently of in the
dominating conduction mechanism and magnetic state of the system. If
the observed $1/f$ noise is due to current independent resistivity
fluctuations which are only probed by the current flow (equilibrium
$1/f$ noise) then the PSD should scale as $S_V=A(T)I^2$, where
$A(T)$ is a temperature dependent constant. Data in Fig. 2 obey the
above relation at all temperatures with the exception of the 77 K.
Noise intensity at 77 K initially increases as $I^2$ but around 1 mA
starts to decrease. The decreasing character is temporarily
suspended by a local noise increase around 2.75 mA and, eventually,
at high currents above 4 mA the noise intensity $S_V(I)$ starts to
increase again. The increase at high currents however, is
accompanied by significantly higher PSD exponent $\alpha$. Joule
heating by the current flow has been excluded as a possible origin
of the noise behavior at low temperatures by independent
measurements of the resistance and magnetic susceptibility at 77 K
as a function of bias current. Noticeable heating effects appeared
only at bias currents exceeding 10 mA.

\begin{figure}
\includegraphics*[width=7truecm]{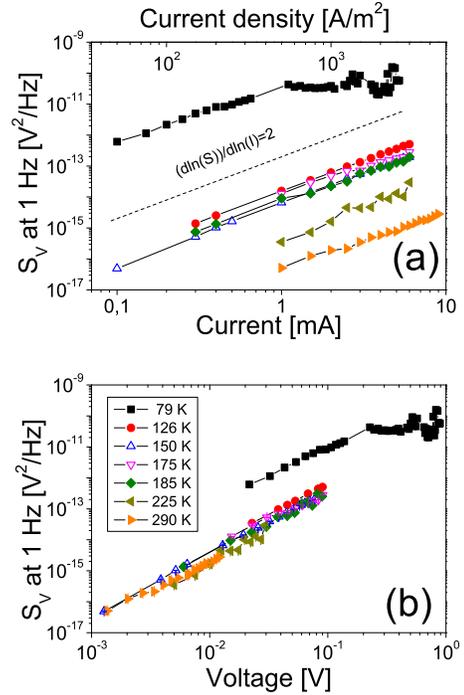}
\caption{ (a) Current dependence of the noise intensity at $f=1$ Hz.
The dashed line is drawn according to $S_V\propto I^2$. (b) Noise
intensity as a function of voltage bias.} \label{SV(I)}
\end{figure}

One can collapse the data from Fig. 2a, again with the exception of
77 K points, to a single line by plotting $S_V(1 Hz)$ as a function
of bias voltage, see Fig. 2b. The collapsed data fit the power law
$S_V=cV^2$ with $c=(4.1 \pm 0.16)\times10^{-11}$ Hz$^{-1}$. This
fact sheds some light on the fluctuation mechanism. $S_V\propto V^2$
means that the normalized $S_v=S_V/V^2=S_R/R^2=S_r=const$, where PSD
of resistance fluctuations, $S_R$, is current and temperature
independent. Since $R$ varies strongly with temperature, by more
than two orders of magnitude, $S_r=const$ means that at all
temperatures the resistivity fluctuation equals to a fixed fraction
of the resistance, $\delta R \propto R$. This is quite a surprising
result considering remarkable changes occurring in the magnetic
state and dissipation mechanism of the sample with changing
temperature. Nevertheless, the fact that current flow can actually
depress the noise at low temperatures is even more astonishing.

Adopting a point of view that low temperature transport in
La$_{0.82}$Ca$_{0.18}$MnO$_{3}$ crystals is dominated by intrinsic
tunneling we plot the intensity of the noise at 77 K together with
the derivative of the conductance as a function of bias voltage in
Fig. 3. We find correlations between behavior of the noise and
$d^2I/dV^2$, which is known to be related to the density of states
involved in the tunneling process. At low bias the noise increases,
as the square of the voltage, until the first peak of $d^2I/dV^2$ at
$V=0.17$ V. Above 0.17 V the noise does not increase with bias and
even decreases in a certain bias range. The noise behavior can be
related to underlining fine structure in $d^2I/dV^2$. However, above
the pronounced $d^2I/dV^2$ peak at $\sim 0.7$ V the noise starts to
increase again.

In general, a peak in $d^2I/dV^2$ is a signature of a step-like
conductivity increase. One can therefore attribute $d^2I/dV^2$ peaks
to openings of additional tunneling channels with higher
conductivity, as predicted by theoretical models of indirect
nonelastic tunneling through a chain of localized
states.\cite{GM,beasley} At low bias voltages the direct and
resonant tunneling mechanism dominate. At higher voltages the
probability of indirect inelastic tunneling is much higher than that
of direct tunneling. The number of involved localized states $N$
increases with increasing bias or with increasing barrier
width.\cite{GM,beasley} The conductivity of an inelastic channel
increases with increasing $N$ and is exponentially higher than the
conductivity of a direct tunneling channel. Eventually, at very high
bias levels, or in very wide barriers, the variable range hopping
(VRH) becomes more favorable than inelastic tunneling
processes.\cite{beasley} Let us underline that in the investigated
system the force exercised by electric field of the bias can
directly influence the topology of PS by stretching insulating FM
phase and thus increasing the effective width of intrinsic tunnel
barriers.\cite{viret}

\begin{figure}
\includegraphics*[width=7truecm]{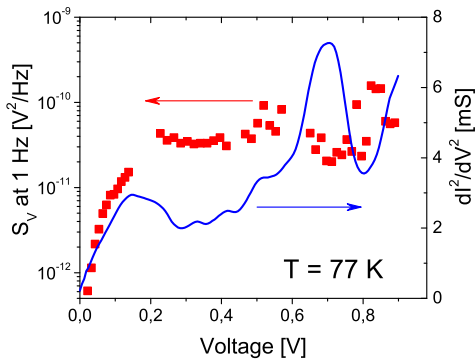}
\caption{ Noise intensity and $d^2I/dV^2$ as a function of bias
voltage.} \label{SVd2IdV2}
\end{figure}

Within this scenario the physical mechanism responsible for decrease
of the noise with increasing current can be the following: In solid
state systems, and in particular in metals and tunnel junctions,
$1/f$ noise is known to result from incoherent superposition of many
elementary two-level fluctuators (TLF) associated with defects and
charge traps. Each TLF generates an elementary random telegraph
signal (RTS) with Lorentzian spectrum. The resulting noise will have
$1/f$ spectrum in the frequency range in which the distribution of
cut-off frequencies within an ensemble is $D(f_c)\propto 1/f_c$. For
thermally activated TLF the latter condition is equivalent to a
requirement of a flat distribution of activation energies. The
maximum power is dissipated by a symmetric TLF, i.e., by a
fluctuator for which the average lifetimes in both RTS levels are
equal. Whenever a TLF becomes asymmetric, the intensity of its
Lorentzian spectrum is depressed. As a result, $1/f$ noise decreases
with increasing asymmetry of elementary fluctuators.

For elementary fluctuators associated with charge traps within a
tunnel barrier, the RTS levels can be identified with empty and
loaded states of the trap. Electrostatic field of the charge trap
modulates the height of the tunnel energy barrier, and consequently
the resistivity of the junction, according to the trap occupancy
state.

Let us assume that all involved traps are symmetric in their
pristine state. It is easy to imagine that electric field of the
applied bias stresses and tilts energy structures of trap-based TLFs
and renders them asymmetric. $1/f$ noise will decrease with
increasing TLF asymmetry due to increasing bias when tunneling
current flows across a channel containing stressed traps.

We tentatively interpret the first $d^2I/dV^2$ peak with a
transition from a direct to indirect tunneling. Observe that in the
elastic tunneling process charge carriers loading the TLF traps are
separated from those constituting the tunneling current. This is not
necessarily the case in inelastic tunneling.

Opening of an inelastic tunneling channel with higher $N$  is marked
by a peak in $d^2I/dV^2$. Local noise maxima at the same bias may be
attributed to partition-like noise associated with additional degree
of freedom: a possibility of conducting current through competing
alternative channels.  When the voltage becomes sufficiently high to
activate the VRH the noise again increases with bias.  Indeed, $S_V
\propto V^2$ is predicted by the model of $1/f$ noise in VRH
regime.\cite{shklovski}

Obviously, in a difference to a discrete fabricated tunnel junction,
or to a single grain boundary junction, \cite{gross} one cannot
provide an absolute direct proof of the tunnel character of the bulk
conductivity dominated by distributed intrinsic junctions. Only an
indirect evidence can be brought from observations of nonlinear
voltage-current $(V-I)$ characteristics and their temperature
evolution, provided both can be well fitted to a tunneling model. In
our previous paper on metastable nonlinear resistivity in $x = 0.18$
LCMO crystals we have well fitted the experimentally observed $V-I$
characteristics to the indirect tunneling model.\cite{prb018}

In conclusion, we have observed equilibrium $1/f$ voltage noise in a
wide range of currents and temperatures corresponding to markedly
different magnetic and transport properties of the system. At low
temperatures where transport is dominated by tunneling mechanism the
nonequilibrium noise appears and decreases with increasing bias due
to bias imposed stress of elementary TLFs participating in inelastic
tunneling.

This work was supported by Israel-Korea bilateral program. X.D.W.
acknowledges support of Monash University "Taft Fellowship" and M.B. of the
Russian-Ukrainian research program "Nanophysics and
Nanoelectronics".

\end{document}